\documentclass[manuscript,screen]{acmart}

\AtBeginDocument{%
  \providecommand\BibTeX{{%
    \normalfont B\kern-0.5em{\scshape i\kern-0.25em b}\kern-0.8em\TeX}}}

\setcopyright{acmcopyright}
\copyrightyear{2021}
\acmYear{2021}
\acmDOI{10.1145/1122445.1122456}

\acmConference[Woodstock '18]{Woodstock '18: ACM Symposium on Neural
  Gaze Detection}{June 03--05, 2018}{Woodstock, NY}
\acmBooktitle{Woodstock '18: ACM Symposium on Neural Gaze Detection,
  June 03--05, 2018, Woodstock, NY}
\acmPrice{15.00}
\acmISBN{978-1-4503-9999-9/18/06}

\newcommand{\Mat}[1]{\bm{#1}}
\newcommand{\Vector}[1]{\bm{#1}}
\newcommand{\Set}[1]{\mathcal{#1}}
\newcommand{\Real}{\mathbb{R}}

\newcommand{\Loss}{\mathcal{L}}

\newcommand{\USet}{\mathcal{U}}
\newcommand{\ISet}{\mathcal{I}}
\newcommand{\SSet}{\mathcal{S}}
\newcommand{\RSet}{\mathcal{R}}

\newcommand{\MLP}{\mathrm{MLP}}
\newcommand{\DIN}{\mathrm{DIN}}
\newcommand{\GRU}{\mathrm{GRU}}
\newcommand{\ReLU}{\mathrm{ReLU}}
\newcommand{\Softmax}{\mathrm{Softmax}}

\usepackage{ulem}
\usepackage{algorithm}
\usepackage{algorithmic}
\usepackage{textcomp}
\usepackage{xcolor}
\usepackage{booktabs}
\usepackage{array}
\usepackage{multirow}
\usepackage{multicol}
\usepackage{enumitem}
\usepackage{subfig}
\usepackage{bm}
\usepackage{url}

\begin{document}

\title{MC$^2$-SF: Slow-Fast Learning for Mobile-Cloud Collaborative Recommendation}


\author{Zeyuan Chen}
\authornotemark[2]
\affiliation{%
  \institution{East China Normal University}
}
\email{chenzyfm@outlook.com}

\author{Jiangchao Yao}
\authornotemark[3]
\affiliation{%
  \institution{Damo Academy, Alibaba Group}
}
\email{jiangchao.yjc@alibaba-inc.com}

\author{Feng Wang}
\authornotemark[3]
\affiliation{%
  \institution{Damo Academy, Alibaba Group}
}
\email{wf135777@alibaba-inc.com}

\author{Kunyang Jia}
\authornotemark[3]
\affiliation{%
 \institution{Damo Academy, Alibaba Group}
}
\email{kunyang.jky@alibaba-inc.com}

\author{Bo Han}
\authornotemark[4]
\affiliation{%
  \institution{Hong Kong Baptist University}
}
\email{bhanml@comp.hkbu.edu.hk}

\author{Wei Zhang}
\authornotemark[2]
\affiliation{%
  \institution{East China Normal University}
}
\email{zhangwei.thu2011@gmail.com}

\author{Hongxia Yang}
\authornotemark[3]
\affiliation{%
  \institution{Damo Academy, Alibaba Group}
}
\email{yang.yhx@alibaba-inc.com}

\renewcommand{\shortauthors}{Trovato and Tobin, et al.}

\begin{abstract}
With the hardware development of mobile devices, it is possible to build the recommendation models on the mobile side to utilize the fine-grained features and \mbox{the real-time feedbacks.} Compared to the straightforward mobile-based modeling appended to the cloud-based modeling, we propose a \underline{S}low-\underline{F}ast learning mechanism to make the \underline{M}obile-\underline{C}loud \underline{C}ollaborative recommendation (MC$^2$-SF) mutual benefit. Specially, in our MC$^2$-SF, the cloud-based model and the mobile-based model are respectively treated as the \emph{slow component} and the \emph{fast component}, according to their interaction frequency in real-world scenarios. During training and serving, they will communicate the prior/privileged knowledge to each other to help better capture the user interests about the candidates, resembling the role of System I and System II in the human cognition. We conduct the extensive experiments on three benchmark datasets and demonstrate the proposed MC$^2$-SF outperforms several state-of-the-art methods. 
\end{abstract}



\maketitle

\section{Introduction}\label{sec:intro}
The information explosion on the websites greatly drives the development of recommender systems, which automatically search the content \textit{e.g.}, movies, songs and news, for the users based on their interests. In the past years, recommender systems are usually deployed in the cloud server, owing to the large amounts of the user behavior data and the high demand of the computing power. Recently, the rapid development of mobile devices reshapes the architecture of industrial recommender systems, and building a recommendation phase on the mobile devices to utilize the fine-grained features and the real-time feedbacks is becoming a trend~\cite{gong2020edgerec,sun2020generic,yao2021device}. 

\begin{figure}[!t]
    \centering
	\includegraphics[width=1\linewidth]{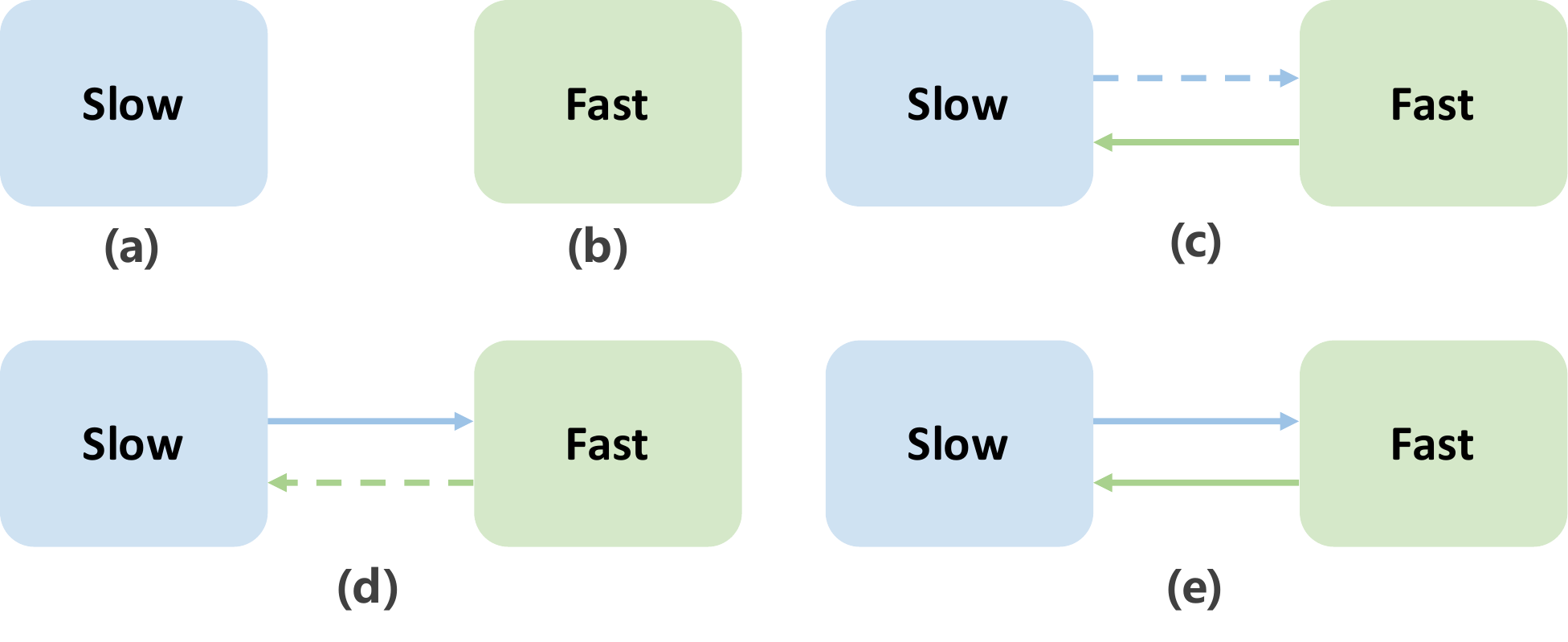}
    \caption{The recommendation prototypes. Subfigure (a) and subfigure (b) represent the independent modeling without relying on other sides, subfigure (c) and subfigure (d) respectively represent the slow-centralized modeling and the fast-centralized modeling and subfigure (e) is our framework.}
    \label{fig:sketch}
\end{figure}

Previous recommendation tends to build the model on the cloud or device side independently as illustrated in Figure~\ref{fig:sketch}(a) and (b). For the slow component deployed in the cloud, it enjoys the large computing power and the rich but delayed user behaviors, which drive the development of a range of deep-learning-based models like SASRec~\cite{KangM18}, DIN~\cite{zhou2018deep} and other complex models~\cite{sun2019bert4rec,li2020time}. 
Regarding the fast component on the mobile side, models are efficient and lightweight to meet the hardware constraint. They usually benefit from the real-time feedbacks, fine-grained features or frequent responses~\cite{gong2020edgerec,sun2020generic} compared to the cloud-based models. However, the models of two sides have no collaboration in the training.

Recent advances in recommendation start to consider the advantages of the counterpart side. For example, one representative  methodology is Federated recommender systems~\cite{yang2020federated}, which leverage the local devices to compute the gradients of the model and simultaneously keep the data privacy. They actually utilize the computing power of the fast component to serve the slow component, but have not considered the modeling of the fast component. We use  Figure~\ref{fig:sketch} (c) to indicate this type of biased collaboration and term it as the \emph{slow-centralized} modeling. Note that, the dashed arrow means the pseudo collaboration for the other side. Another representative methodology is Model-Personalized recommender systems~\cite{yao2021device}, which leverage the cloud server and data to re-calibrate the backbone model. We can appropriately consider it as the reverse counterpart of Federated recommender systems and illustrate it in Figure~\ref{fig:sketch} (d). In comparison, we term it as the \emph{fast-centralized} modeling.

Different from the above works, we focus on the bidirectional collaboration to benefit both the slow component and the fast component as shown in Figure~\ref{fig:sketch} (e). Specifically, we propose MC$^2$-SF, a Slow-Fast Learning mechanism for Mobile-Cloud Collaborative recommendation. In MC$^2$-SF, the slow component helps the fast component make predictions by delivering the auxiliary latent representations; and conversely, the fast component transfers the feedbacks from the real-time exposed items to the slow component, which helps better capture the user interests. The intuition behind MC$^2$-SF resembles the role of System I and System II in the human recognition~\cite{kahneman2011thinking}, where System II makes the slow changes but conducts the comprehensive reasoning along with the circumstances, and System I perceives fast to make the accurate recognition~\cite{madan2021fast}. The interaction between System I and System II allows the prior/privileged information exchanged in time to collaboratively meet the 
requirements of the environment. 

We summarize the contributions of this paper as follows:
\begin{itemize}[leftmargin=*]
    \item To our best knowledge, we are the first to study the bidirectional collaboration between the cloud-based model and the mobile-based model with the hardware advances.
    
    \item We introduce a slow-fast learning mechanism, MC$^2$-SF, which considers the cloud-based model and the mobile-based model respectively as a slow component and a fast component, between which the prior and the privileged knowledge are interacted during the training and serving.
    
    \item Extensive experiments on three benchmark datasets have demonstrated that the proposed method significantly outperforms the state-of-the-art recommendation baselines and shown the promise of the bidirectional collaboration.
    
\end{itemize} 

\section{Related Work}\label{sec:related}

\subsection{Independent Modeling in Recommendation}
In this section, we review the independent modeling on the cloud side and on the mobile side respectively. Regarding the cloud-based recommendation models, the early exploration falls in the collaborative filtering~\cite{wang2006unifying,su2009survey} or Latent Factor Model (LFM)~\cite{zhang2013combining}. With the development of deep learning, deep neural networks are involved into the recommendation to acquire the high-level semantics~\citep{elkahky2015multi,van2013deep,wang2015collaborative}. For example, DeepFM~\cite{guo2017deepfm} combines the power of the factorization machines for recommendation and deep learning for feature learning, yielding a promising improvement in performance. To model the user dynamics in the sequences, the recurrent neural networks (RNNs) are applied to obtain the representations of whole user behavior sessions for sequential recommendation~\cite{hidasi2015session}. The subsequent work~\cite{quadrana2017personalizing} further investigates the hierarchical version of recurrent neural networks.
Moreover, Caser~\cite{TangW18} considers to capture the local features in sequential behaviors by means of the convolution filters. Some recent studies explore to utilize the attention mechanisms to learn behavior sequence representations~\cite{VaswaniSPUJGKP17,KangM18,zhou2018deep}, or apply sequential recommendation to the specific scenarios~\cite{LianWG0C20,chen2021dual,zhang2021cause}.

With the hardware development of mobile devices, the mobile-based modeling has drawn much more attention in the industrial scenarios~\cite{satyanarayanan2017emergence}. For example, knowledge distillation has been widely used to distill a smaller but efficient student network from the teacher network, which adapts the on-device inference~\cite{kim2016sequence,zhou2018rocket}. The compression technique that jointly leverages weight quantization and distillation for efficiently executing deep models in resource-constrained environments like mobile or embedded devices, is also proposed. CpRec~\cite{sun2020generic} employs a generic model shrinking technique to reduce responding time and memory footprint. Other work~\cite{gong2020edgerec}
implements a novel recommender system on the device side and addresses the serving concern based on a split deployment strategy.

\subsection{Biased Collaboration in Recommendation}
This line of works could be split into two parts, the slow-centralized modeling and the fast-centralized modeling. One exemplar of the slow-centralized modeling is Federated recommender systems ~\cite{zhou2012federated}, which trains the cloud-based model with the aid of distributed local devices. The gradients of the local copy from the centralized deep model are first executed on plenty of devices, and then are collected to the server to update the model parameters by federated averaging (FedAvg) or its variants ~\cite{anelli2020federank,muhammad2020fedfast}. For example, Qi \textit{et al.,}~\cite{qi2020privacy} leveraged the local information of massive users to train an accurate news recommendation model and meanwhile keep the privacy of the sensitive data. To reduce the communication costs of federated learning, the structured update and the sketched update to compress networks on the device side are introduced~\cite{konevcny2016federated}. One representative work of the fast-centralized modeling is DCCL~\cite{yao2021device}, which could be treated as the reverse counterpart of Federated Learning. DCCL leverages the cloud server and data to re-calibrate the backbone model, but actually for the on-device personalization. The slow-centralized modeling and the fast-centralized modeling do not actually achieve the bidirectional mobile-cloud collaboration, which is meaningful under the real-world pipeline of the industrial recommender systems.

\section{Preliminary}\label{sec:pre}
In this section, we will formulate the problem of Mobile-Cloud collaborative recommendation. Let $\USet=\{u_1,...,u_N\}$ denote the user set and $\ISet=\{i_1,...,i_M\}$ denote the item set, where $N$ and $M$ are the user number and the item number respectively. For each user $u_n$, we define the interactive item sequence $\SSet_{u_n}=\{i_1,...,i_{l}\}$. In the perspective of collaborative recommendation, $\SSet_{u_n}$ on the cloud side is instantiated as  $\SSet_{u_n}^S=\{i_1^S,i_2^S,...,i_{l^S}^S\}$ and on the mobile side is instantiated as $\SSet_{u_n}^F=\{i_1^F,i_2^F,...,i_{l^F}^F\}$ respectively. The reason that the sequences on two sides are different, is that the feature on the mobile side are more real-time and fine-grained, and we might not leverage too long sequences \textit{i.e.,} $l^F<l^S$, given the limited computational resource of mobile devices. We will refer more details about them in the subsequent sections. Without loss of generality, we define the problem of mobile-cloud collaborative recommendation as follows.

\newtheorem*{problem}{Problem}
\begin{problem}[Mobile-Cloud Collaborative Recommendation] As aforementioned before, we term the cloud-based model as the slow component and the mobile-based model as the fast component.  Given a target user $u$ and a candidate item $i$, the goal of both slow and fast components is to learn a function that accurately predicts the user interaction, respectively defining as $\hat{y}^S=f^S(u,i|\SSet_{u_n}^S,\RSet_{u_n}^F;\Theta^{S})$ and $\hat{y}^F=f^F(u,i|\SSet_{u_n}^F,\RSet_{u_n}^S;\Theta^{F})$.
$\Theta^{S}$ and $\Theta^{F}$ denotes the trainable parameters of the slow component and the fast component, and $\RSet_{u_n}^F$ and $\RSet_{u_n}^S$ mean the interactive features between them. Compared to the traditional recommendation without mobile-cloud collaboration or only with the biased collaboration, they will not have $\RSet_{u_n}^F$ and $\RSet_{u_n}^S$. The Figure~\ref{fig:sketch} summarizes the difference from the previous works.
\end{problem}

\begin{figure*}[!t]
    \centering
	\includegraphics[width=.86\linewidth]{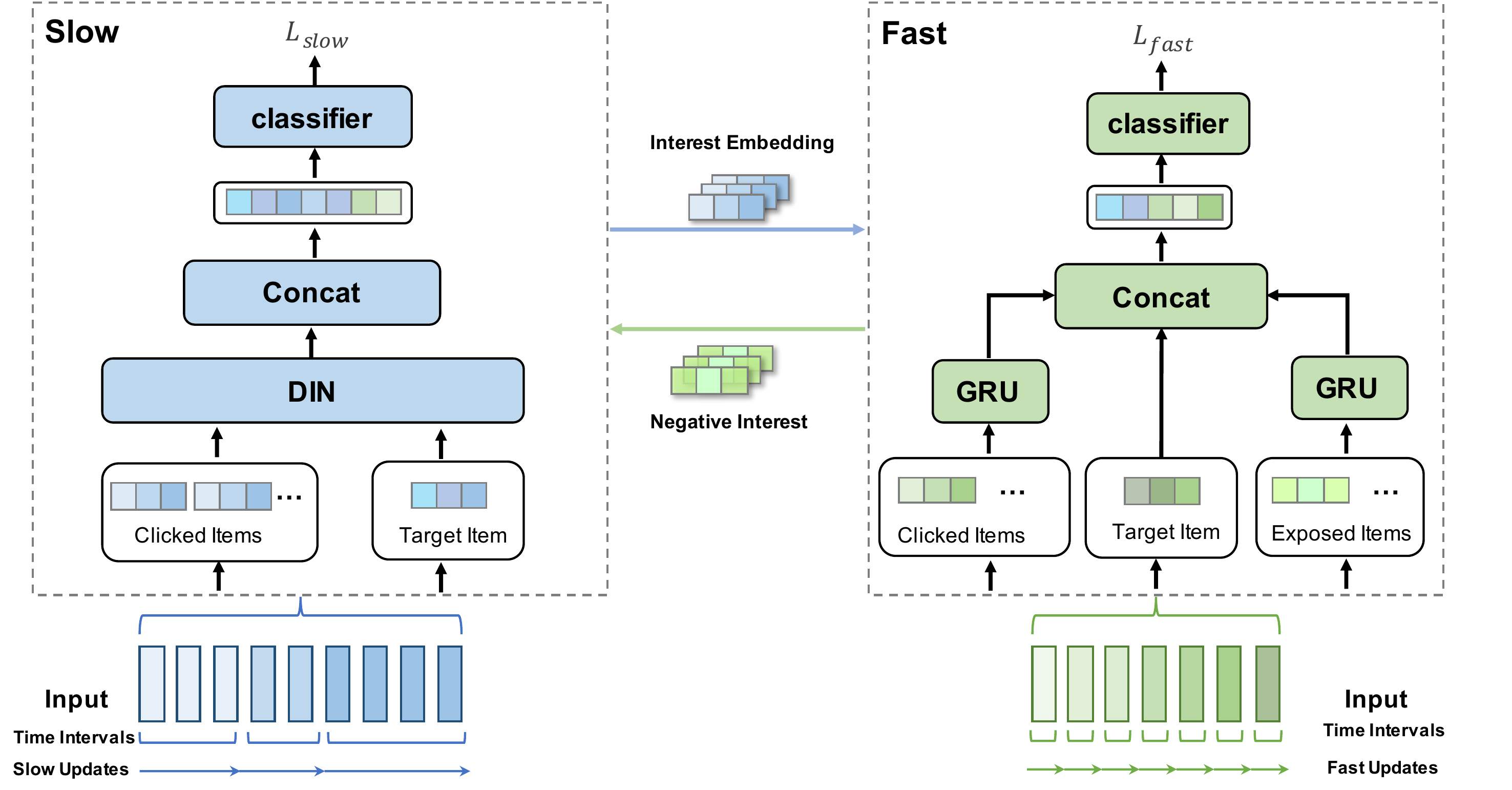}
    \caption{Architecture of the model MC$^2$-SF. Input denotes the sequence of item embedding. The interacted items generated in fast component would be updated instantly (fast update) and uploaded to slow component at regular intervals (slow update) in industrial practice, whose input update is slower than fast component. Owing to large computing power, slow component has more parameters and computational complexity compared to fast component. The collaboration of both sides is achieved by prior/privileged knowledge delivery. }
    \label{fig:model-architecture}
\end{figure*}

\section{Methodology}

Following the convention of recent deep learning-based recommendation methods, we map each item ID to a dense vector and tune it in the training stage. Taking item $i$ as an example, we define the following embedding-lookup function:
\begin{equation}\label{eq:map} 
 \Vector{e}_i = \Mat{E}_I\Vector{o}_i\,, \nonumber
\end{equation}
where $\Mat{E}_I\in\Real^{d\times N}$ is a trainable item embedding matrix and $d$ is the embedding dimension. Noting that, $\Mat{E}_I$ is saved in the cloud and only part of it is distributed to the mobile when requested.
$\Vector{o}_i$ is an one-hot vector for the $i$-th item.

\subsection{Independent Slow Component}
For the cloud-based modeling, \textit{i.e.,} the slow component, we use  $\mathcal{H}^S=[\Vector{e}_1^S,\Vector{e_2}^S,...,\Vector{e}_{l^S}^S]$ to represent the representation of the click sequence $\SSet_{u}^S$. Deep Interest Network~\cite{zhou2018deep}, abbreviated as $\DIN$, is used to model the user representation from $\mathcal{H}^S$ and the candidate item $\Vector{e}_i$ as follows,
\begin{equation}\label{eq:din}
    \Vector{r}_t = \DIN(\mathcal{H}^S,~\Vector{e}_i),
\end{equation}
where $\Vector{e}_i$ denotes the embedding of the candidate item. We then combine $\Vector{r}_t$ with $\Vector{e}_i$ and feed into $\MLP$ to compute the following click-through rate prediction
\begin{equation}
    \hat{y}^S = \sigma(\MLP(\Vector{r}_t\oplus \Vector{e}_i)). \nonumber
\end{equation}
Finally, the cross-entropy loss is applied for the training of the parameters in the slow component as follows,
\begin{equation}\label{eq:loss_slow}
    \Loss^S = -\big(y^{S}\log(\hat{y}^S)+(1-y^{S})\log(1-\hat{y}^S)\big)\,.
\end{equation} 
By far, we present a typical cloud-based model without any real-time knowledge intervention from the mobile side.

\subsection{Independent Fast Component}
For the mobile-based model, \textit{i.e.,} the fast component, we use $\mathcal{H}^F=[\Vector{e}_1^F,\Vector{e}_2^F,...,\Vector{e}_{l^F}^F]$ to denote the representations of either clicked or real-time exposed items $\SSet_{u}^F$. Note that, the reason that we take the exposed items into account, is that they are also informative as pointed in~\cite{gong2020edgerec,xie2020deep}. 
In the fast component, we use two GRUs to separately encode the clicked parts and the exposed parts of $\mathcal{H}^F$, termed as $\mathcal{H}^F_p$ and $\mathcal{H}^F_n$ respectively, and then fuse them by a target-aware attention mechanism. Formally, the procedure is formulated by the following equations,
\begin{align}\label{eq:GRU_fusion}
    \begin{split}
        & \Vector{r}_1 = \GRU_p(\mathcal{H}^F_p), ~~\Vector{r}_2 = \GRU_n(\mathcal{H}^F_n), \\
        & \Vector{r} = \left[ \begin{array}{c}\Vector{r}_1 \\ \Vector{r}_2\end{array}\right]^\top \Softmax\left(\left[\begin{array}{c}\MLP(\Vector{r}_1\oplus \Vector{e}_i) \\ \MLP(\Vector{r}_2\oplus \Vector{e}_i)\end{array}\right]\right), 
    \end{split}
\end{align}
where $\Vector{r}_1$ can be seen as the user real-time interests, and $\Vector{r}_2$ can be considered as the negative impression. Like the slow component, we combine $\Vector{r}$ and $\Vector{e}_i$ to compute the prediction
\begin{equation}
    \hat{y}^F = \sigma(\MLP(\Vector{r}\oplus \Vector{e}_i)). \nonumber
\end{equation}
Similarly, the cross-entropy loss is applied to train the parameters of the fast component as follows,
\begin{equation}\label{eq:loss_fast}
    \Loss^F = -\big(y^{F}\log(\hat{y}^F)+(1-y^{F})\log(1-\hat{y}^F)\big)\,.
\end{equation}
Now, we have an independent recommendation model on the mobile side, which exploits more real-time clicked items and feedbacks from the exposed items to capture user interests.

\subsection{Interactive Slow Component}
As proved by~\cite{gu2021real}, the real-time exposures reflecting the user negative impression would bring the gains to the model.
Based on Eq.~\eqref{eq:GRU_fusion} in the basic fast component, we can obtain the negative memory $\hat{\Vector{r}}_2$ which memorizes the previous exposed items, and send it to the slow component. With this, the slow component can continue to feed $\Vector{e}_{i}$ into $\GRU_n$ to generate the presumed response, and re-weight it to form the final feature about the exposure feedback,
\begin{align} \label{eq:int_slow_gru}
\begin{split}
    & \Bar{\Vector{r}}_2 = \GRU_n(\Vector{e}_{i})\big|_{\hat{\Vector{r}}_2}  \\
    & \Vector{r}_n = \Bar{\Vector{r}}_2 * \sigma(\Vector{W}_S[ \Bar{\Vector{r}}_2\oplus \Vector{e}_i]+b_S).
\end{split}
\end{align}
Note that, during the training phase, the offline optimization of the slow component could affect the update operation of $\GRU_n$.
After Eq.~\eqref{eq:int_slow_gru}, we can concatenate $\Vector{r}_n$, $\Vector{r}_t$ in Eq.~\eqref{eq:din} and $\Vector{e}_i$ to compute the exposure-aware prediction
\begin{align} \label{eq:int_slow_logit}
\begin{split}
    & \hat{y}^S = \sigma(\MLP(\Vector{r}_n\oplus \Vector{r}_t \oplus \Vector{e}_i)).
\end{split}
\end{align}
The optimization objective is the same as Eq.~\eqref{eq:loss_slow}.

\subsection{Interactive Fast Component}
The fast component is usually a smaller model than the slow component, since the cloud side has the sufficient computing and storage resource to enlarge the model. This yields that the prior knowledge from the slow component is a useful hint to the fast component. Considering this, we send $\Vector{r}_n$ and $\Vector{r}_t$ from the cloud to the mobile to help the prediction of the fast component. Specially, we transform them into the space of the hidden state in GRU, and use the transformed representations to initialize the state of $\GRU_p$ and $\GRU_n$ in Eq.~\eqref{eq:GRU_fusion}. The procedure is formulated as follows,
\begin{align}
\begin{split}
    & \Vector{r}_f = \alpha\Vector{r}_n+(1-\alpha)\Vector{r}_t,
    \alpha = \sigma\left(\Vector{W}^0_F[\Vector{r}_n\oplus \Vector{r}_t]+b^0_F\right),\\
    &  \Vector{r}_p = \GRU_p(\mathcal{H}^F_p)\big|_{\Vector{z}_p=\ReLU(\Vector{W}^1_F\Vector{r}_f+b^1)}, \\ 
    & \Tilde{\Vector{r}}_n = \GRU_n(\mathcal{H}^F_n)\big|_{\Vector{z}_n = \ReLU(\Vector{W}^2_F\Vector{r}_f+b^2)}~.
\end{split}
\end{align}
Following Eq.~\eqref{eq:GRU_fusion}, we could transform $\Vector{r}_p$ and $\Tilde{\Vector{r}}_n$ to the final representation $\Vector{r}$. To enhance the effect of the prior knowledge from the slow component, we also combine $r_t$ with $\Vector{r}$ and $\Vector{e}_i$ to compute the prediction score like Eq.~\eqref{eq:int_slow_logit} as follows
\begin{equation}
    \hat{y}^F = \sigma(\MLP(\Vector{r}\oplus \Vector{r}_t \oplus \Vector{e}_i))\,.
\end{equation}
Similarly, the optimization objective Eq.~\eqref{eq:loss_fast} is applied. 

\subsection{Slow-Fast Learning}
The complete procedure of MC$^2$-SF is summarized in Algorithm~\ref{alg:algorithm}. Specifically, the slow component firstly receives negative memory $\hat{\Vector{r}}_2$ and proceeds model optimization based on $\hat{\Vector{r}}_2$ and relevant input. With the completion of the training, the slow could generate representation $\Vector{r}_n$ and $\Vector{r}_t$ so as to send them to the fast component to help prediction, which corresponds to ``Interest Embedding'' transfer operation as shown in Figure~\ref{fig:model-architecture}. Based on $\Vector{r}_n$ and $\Vector{r}_t$, the fast component continues to optimize. As the model is deployed in the mobile, it could update corresponding real-time exposed items to generate new negative memory $\hat{\Vector{r}}_2$. The upload of ``Negative Interest'' in Figure~\ref{fig:model-architecture} represents new negative memory $\hat{\Vector{r}}_2$ is uploaded to improve the slow component. As such, the collaboration framework could be iterate continuously.

\begin{algorithm}[tb]
\caption{Slow-Fast Learning}
\label{alg:algorithm}
\begin{algorithmic} 
\WHILE{lifecycle}
\STATE$\textbf{Slow:}$\\
\quad $\textbf{Training:}$\\
\quad\quad 1.Receive negative memory $\hat{\Vector{r}}_2$\\
\quad\quad 2.Optimize the slow component $f^S$\\
\quad $\textbf{Inference:}$\\
\quad\quad 1.Generate representations $\Vector{r}_n$ and $\Vector{r}_t$ based on $f^S$.\\
\quad\quad 2.Send representations to the fast component $f^F$.\\
$\textbf{Fast:}$\\
\quad $\textbf{Training:}$\\
\quad\quad 1.Receive representations $\Vector{r}_n$ and $\Vector{r}_t$\\
\quad\quad 2.Optimize the fast component $f^F$\\
\quad $\textbf{Inference:}$\\
\quad\quad 1.Accumulate exposed items and update negative memory $\hat{\Vector{r}}_2$\\
\quad\quad 2.If time $\textgreater$ threshold: upload $\hat{\Vector{r}}_2$ \\
\quad\quad 3.Else: return step 1\\
\ENDWHILE
\end{algorithmic}
\end{algorithm}

\section{Experiments}\label{sec:exp}
This section first clarifies the experimental setups and then
provides comprehensive experimental results, striving to answer
the pivotal questions below:

\begin{itemize}[leftmargin=1.8em]
\item[\textbf{\texttt{Q1}.}] What is the performance of MC$^2$-SF compared with the state-of-the-art recommendation models?

\item[\textbf{\texttt{Q2}.}] How does the bidirectional collaboration used in MC$^2$-SF contribute to the overall performance?

\item[\textbf{\texttt{Q3}.}] How do the hyper-parameters of MC$^2$-SF make the effect on the final performance?
\end{itemize}

\subsection{Experimental Setups}
This section explains the used datasets, the adopted evaluation
protocols, the baselines to be compared, and the model
implementations.

\begin{table}[!t]
\centering
\caption{Statistics of the datasets.}\label{tbl:stat}
\begin{tabular}{l|ccc}
\hline
\textbf{Dataset}  & ML-1M  & Alipay  & Steam \\ \hline
 \# Users & 6,040 & 1,031,268 & 39,795 \\
 \# Items & 3,706 & 13,932 & 14,411 \\
 \# Interactions & 1,000,209 & 228,412,772
 & 17,978,753 \\
\hline
\end{tabular}
\end{table}

\subsubsection{Datasets} 
To evaluate the model performance, we choose two datasets, ML-1M\footnote{\url{https://files.grouplens.org/datasets/movielens/ml-1m.zip}} and Steam\footnote{\url{http://cseweb.ucsd.edu/~wckang/steam_reviews.json.gz}} that are publicly available and one industrial dataset Alipay. We preprocess the datasets to guarantee that users and items have been interacted at least 20 times in above datasets. The basic statistics of the three datasets are summarized in Table~\ref{tbl:stat}. Generally, each dataset is divided into three disjoint parts according to the log timestamps, including slow training phase, fast training phase, and testing phase. Specifically, for a sequence containing $l$ items, testing phase contains the last 5 items, fast training phase uses the interaction history from the $(l-9)$th to the $(l-5)$th item and slow training phase contains the rest of items. Each component is trained in the corresponding phase with the aid of the other side. 

\subsubsection{Evaluation Protocols}
Three widely used metrics is adopted: (1) HR@k (Hit Ratio@k) is the proportion of recommendation lists that have at least one positive item within top-k positions.
(2) NDCG@k (Normalized Discounted Cumulative Gain@k) is a position-aware ranking metric that assigns larger weights to the top positions.
As the positive items rank higher, the metric value becomes larger.
(3) MRR (Mean Reciprocal Rank) measures the relative position of the top-ranked positive item and takes value 1 if the positive item is ranked at the first position.
HR@k and MRR mainly focus on the first positive item, while NDCG@k considers a wider range of positive items.
They are mathematically defined as follows. 
\begin{align}
    \text{HR@k} &= \frac{1}{|\USet|}\sum_{u\in \Set{U}}\mathbb{I}(P_{u,i}\leq k) \,, \notag \\
    \text{NDCG@k} &= \frac{1}{|\USet|}\sum_{u\in \Set{U}} \frac{2^{\mathbb{I}(P_{u,i}\leq k)}-1}{log(P_{u,i}+1)}\,,  \notag\\
    \text{MRR} &= \frac{1}{|\USet|}\sum_{u\in \Set{U}}\frac{1}{P_{u,i}} \,.\notag
\end{align}
where $P_{u,i}$ is the ranking position of interaction between user $u$ and item $i$, and $\mathbb{I}$ is the indicator function.

In the experiments, we list the results w.r.t. HR@1, HR@5, HR@10, NDCG@5, NDCG@10 and MRR. 

Regarding negative sampling, we follow the way which is commonly
observed in recommendation studies considering implicit
feedback~\cite{HeLZNHC17,KangM18}.
It is notable that there is a difference between our settings and traditional recommendation when testing. For each sequence, traditional recommendation uses the most recent interaction of each user for testing. However, in our settings, we have different strategies to evaluate the slow component and fast component. The slow component cannot immediately use the sequence on the test set so that it predicts every item in testing phase based on the sequence in the slow training phase. But for the fast component deployed on the mobile side, it could instantaneously access the sequence of testing phase. Therefore, the fast component leverages all the items before the target item to perform the prediction.

\begin{table*}[!t]
\centering
\caption{The results w.r.t. HR@1, HR@5, HR@10, NDCG@5, NDCG@10, and MRR for recommendation on three datasets. The best results in each measure are highlighted in “bold”. Improv. denotes the relative improvement over the second-best results.}\label{tbl:performance-comp}
\begin{tabular}{cccccccccccc}
\hline
\multicolumn{1}{c|}{\multirow{3}{*}{Dataset}}&
 \multicolumn{1}{c|}{\multirow{3}{*}{Metric}}&
 \multicolumn{1}{c}{}& 
 \multicolumn{1}{c}{Fast} & 
 \multicolumn{1}{c|}{} & 
 \multicolumn{1}{c|}{Slow} & 
 \multicolumn{3}{c|}{Vanilla Slow+Fast} & 
 \multicolumn{1}{c|}{\multirow{3}{*}{MC$^2$-SF}} &
 \multicolumn{1}{c}{\multirow{3}{*}{Improv.}} \\ 
 \cline{3-9}
\multicolumn{1}{c|}{}&
 \multicolumn{1}{c|}{}&
 \multicolumn{1}{c}{\multirow{2}{*}{FM}}& 
 \multicolumn{1}{c}{\multirow{2}{*}{NeuMF}} & 
 \multicolumn{1}{c|}{\multirow{2}{*}{GRU4Rec}} & 
 \multicolumn{1}{c|}{\multirow{2}{*}{DIN}} & 
 \multicolumn{1}{c}{} & 
 \multicolumn{1}{c}{DIN+} & 
 \multicolumn{1}{c|}{} &
 \multicolumn{1}{c|}{} &
 \multicolumn{1}{c}{}\\ 
 \multicolumn{1}{c|}{}&
 \multicolumn{1}{c|}{}&
 \multicolumn{1}{c}{} & 
 \multicolumn{1}{c}{} & 
 \multicolumn{1}{c|}{} & 
 \multicolumn{1}{c|}{}&
 \multicolumn{1}{c}{FM} & 
 \multicolumn{1}{c}{NeuMF} & 
 \multicolumn{1}{c|}{GRU4Rec} &
 \multicolumn{1}{c|}{} &
  \\ 
\hline
\multicolumn{1}{c|}{\multirow{6}{*}{ML-1M}} &

\multicolumn{1}{c|}{HR@1}    & 
\multicolumn{1}{c}{0.5308}       & 
\multicolumn{1}{c}{0.5345}       & 
\multicolumn{1}{c|}{0.5459}    & 
\multicolumn{1}{c|}{0.5505}    & 
\multicolumn{1}{c}{0.5508}      & 
\multicolumn{1}{c}{0.5441}    & 
\multicolumn{1}{c|}{0.5671}      & 
\multicolumn{1}{c|}{\textbf{0.6105}} &
\multicolumn{1}{c}{7.65\%}      \\

\multicolumn{1}{c|}{} &

\multicolumn{1}{c|}{HR@5}    & 
\multicolumn{1}{c}{0.6619}       & 
\multicolumn{1}{c}{0.6622}       & 
\multicolumn{1}{c|}{0.6794}    & 
\multicolumn{1}{c|}{0.6868}    & 
\multicolumn{1}{c}{0.6782}      & 
\multicolumn{1}{c}{0.6709}    & 
\multicolumn{1}{c|}{0.6880}      & 
\multicolumn{1}{c|}{\textbf{0.7166}}      & 
\multicolumn{1}{c}{4.16\%} \\

\multicolumn{1}{c|}{} &

\multicolumn{1}{c|}{HR@10}   & 
\multicolumn{1}{c}{0.7511}       & 
\multicolumn{1}{c}{0.7538}       & 
\multicolumn{1}{c|}{0.7615}    & 
\multicolumn{1}{c|}{0.7616}    & 
\multicolumn{1}{c}{0.7567}      & 
\multicolumn{1}{c}{0.7618}    & 
\multicolumn{1}{c|}{0.7643}      & 
\multicolumn{1}{c|}{\textbf{0.7768}}      & 
\multicolumn{1}{c}{1.64\%}       \\

\multicolumn{1}{c|}{} &

\multicolumn{1}{c|}{NDCG@5}  & 
\multicolumn{1}{c}{0.6150}       & 
\multicolumn{1}{c}{0.6230}       & 
\multicolumn{1}{c|}{0.6202}    & 
\multicolumn{1}{c|}{0.6207}    & 
\multicolumn{1}{c}{0.6271}      & 
\multicolumn{1}{c}{0.6265}    & 
\multicolumn{1}{c|}{0.6285}      & 
\multicolumn{1}{c|}{\textbf{0.6643}}      & 
\multicolumn{1}{c}{5.70\%}\\

\multicolumn{1}{c|}{} &

\multicolumn{1}{c|}{NDCG@10} & 
\multicolumn{1}{c}{0.6395}       & 
\multicolumn{1}{c}{0.6471}       & 
\multicolumn{1}{c|}{0.6445}    & 
\multicolumn{1}{c|}{0.6449}    & 
\multicolumn{1}{c}{0.6508}      & 
\multicolumn{1}{c}{0.6509}    & 
\multicolumn{1}{c|}{0.6538}      & 
\multicolumn{1}{c|}{\textbf{0.6837}}      & 
\multicolumn{1}{c}{4.57\%}\\

\multicolumn{1}{c|}{} &

\multicolumn{1}{c|}{MRR}     &
\multicolumn{1}{c}{0.6108}       & 
\multicolumn{1}{c}{0.6162}       & 
\multicolumn{1}{c|}{0.6177}    & 
\multicolumn{1}{c|}{0.6197}    & 
\multicolumn{1}{c}{0.6169}      & 
\multicolumn{1}{c}{0.6216}    & 
\multicolumn{1}{c|}{0.6304}      & 
\multicolumn{1}{c|}{\textbf{0.6636}}      &
\multicolumn{1}{c}{5.27\%}\\ 
\hline
\hline
\multicolumn{1}{c|}{\multirow{6}{*}{Alipay}} &

\multicolumn{1}{c|}{HR@1}    &
\multicolumn{1}{c}{0.3462}       & 
\multicolumn{1}{c}{0.2974}       & 
\multicolumn{1}{c|}{0.5756}    & 
\multicolumn{1}{c|}{0.6086}    & 
\multicolumn{1}{c}{0.6043}      & 
\multicolumn{1}{c}{0.5395}    & 
\multicolumn{1}{c|}{0.6224}      & 
\multicolumn{1}{c|}{\textbf{0.6286}}      & 
\multicolumn{1}{c}{1.00\%}\\

\multicolumn{1}{c|}{} &

\multicolumn{1}{c|}{HR@5}    & 
\multicolumn{1}{c}{0.5669}       & 
\multicolumn{1}{c}{0.5130}       & 
\multicolumn{1}{c|}{0.7073}    & 
\multicolumn{1}{c|}{0.7222}    & 
\multicolumn{1}{c}{0.7221}      & 
\multicolumn{1}{c}{0.6763}    & 
\multicolumn{1}{c|}{0.7301}      & 
\multicolumn{1}{c|}{\textbf{0.7456}}      &
\multicolumn{1}{c}{2.12\%}\\

\multicolumn{1}{c|}{} &

\multicolumn{1}{c|}{HR@10}   & 
\multicolumn{1}{c}{0.7407}       & 
\multicolumn{1}{c}{0.6943}       & 
\multicolumn{1}{c|}{0.8346}    & 
\multicolumn{1}{c|}{0.8474}    & 
\multicolumn{1}{c}{0.8473}      & 
\multicolumn{1}{c}{0.8107}    & 
\multicolumn{1}{c|}{0.8545}      & 
\multicolumn{1}{c|}{\textbf{0.8708}}      & 
\multicolumn{1}{c}{1.91\%} \\

\multicolumn{1}{c|}{} &

\multicolumn{1}{c|}{NDCG@5}  & 
\multicolumn{1}{c}{0.4586}       & 
\multicolumn{1}{c}{0.4050}       & 
\multicolumn{1}{c|}{0.6411}    & 
\multicolumn{1}{c|}{0.6590}    & 
\multicolumn{1}{c}{0.6631}      & 
\multicolumn{1}{c}{0.6063}    & 
\multicolumn{1}{c|}{0.6744}      & 
\multicolumn{1}{c|}{\textbf{0.6852}}      & 
\multicolumn{1}{c}{1.60\%}\\

\multicolumn{1}{c|}{} &

\multicolumn{1}{c|}{NDCG@10} & 
\multicolumn{1}{c}{0.5144}       & 
\multicolumn{1}{c}{0.4633}       &
\multicolumn{1}{c|}{0.6820}    & 
\multicolumn{1}{c|}{0.6993}    & 
\multicolumn{1}{c}{0.7032}      & 
\multicolumn{1}{c}{0.6496}    & 
\multicolumn{1}{c|}{0.7124}      & 
\multicolumn{1}{c|}{\textbf{0.7256}}      & 
\multicolumn{1}{c}{1.85\%}\\

\multicolumn{1}{c|}{} &

\multicolumn{1}{c|}{MRR}     &
\multicolumn{1}{c}{0.4623}       & 
\multicolumn{1}{c}{0.4127}       &
\multicolumn{1}{c|}{0.6470}    &
\multicolumn{1}{c|}{0.6640}    & 
\multicolumn{1}{c}{0.6668}      & 
\multicolumn{1}{c}{0.6116}    & 
\multicolumn{1}{c|}{0.6804}      & 
\multicolumn{1}{c|}{\textbf{0.6906}}      & 
\multicolumn{1}{c}{1.50\%}\\ \hline
\hline
\multicolumn{1}{c|}{\multirow{6}{*}{Steam}} &

\multicolumn{1}{c|}{HR@1}    & 
\multicolumn{1}{c}{0.4204}       & 
\multicolumn{1}{c}{0.3310}       & 
\multicolumn{1}{c|}{0.4492}    & 
\multicolumn{1}{c|}{0.4512}    & 
\multicolumn{1}{c}{0.4699}      & 
\multicolumn{1}{c}{0.4150}    & 
\multicolumn{1}{c|}{0.4796}      & 
\multicolumn{1}{c|}{\textbf{0.4926}}      & 
\multicolumn{1}{c}{2.71\%}\\

\multicolumn{1}{c|}{} &

\multicolumn{1}{c|}{HR@5}    & 
\multicolumn{1}{c}{0.5393}       & 
\multicolumn{1}{c}{0.5145}       &
\multicolumn{1}{c|}{0.5398}    & 
\multicolumn{1}{c|}{0.5425}    & 
\multicolumn{1}{c}{0.5679}      & 
\multicolumn{1}{c}{0.5281}    & 
\multicolumn{1}{c|}{0.5726}      & 
\multicolumn{1}{c|}{\textbf{0.5853}}      & 
\multicolumn{1}{c}{2.22\%}\\

\multicolumn{1}{c|}{} &

\multicolumn{1}{c|}{HR@10}   & 
\multicolumn{1}{c}{0.6054}       & 
\multicolumn{1}{c}{0.5729}       &
\multicolumn{1}{c|}{0.6111}    & 
\multicolumn{1}{c|}{0.6098}    & 
\multicolumn{1}{c}{0.6256}      & 
\multicolumn{1}{c}{0.5989}    & 
\multicolumn{1}{c|}{0.6419}      & 
\multicolumn{1}{c|}{\textbf{0.6538}}      & 
\multicolumn{1}{c}{1.85\%}\\

\multicolumn{1}{c|}{} &

\multicolumn{1}{c|}{NDCG@5}  & 
\multicolumn{1}{c}{0.4821}       & 
\multicolumn{1}{c}{0.4372}       & 
\multicolumn{1}{c|}{0.4950}    & 
\multicolumn{1}{c|}{0.4976}    & 
\multicolumn{1}{c}{0.5202}      & 
\multicolumn{1}{c}{0.4666}    & 
\multicolumn{1}{c|}{0.5265}      & 
\multicolumn{1}{c|}{\textbf{0.5389}}      & 
\multicolumn{1}{c}{2.30\%}\\

\multicolumn{1}{c|}{} &

\multicolumn{1}{c|}{NDCG@10} & 
\multicolumn{1}{c}{0.5034}       & 
\multicolumn{1}{c}{0.4600}       &
\multicolumn{1}{c|}{0.5178}    & 
\multicolumn{1}{c|}{0.5192}    & 
\multicolumn{1}{c}{0.5388}      & 
\multicolumn{1}{c}{0.4854}    & 
\multicolumn{1}{c|}{0.5487}      & 
\multicolumn{1}{c|}{\textbf{0.5617}}      & 
\multicolumn{1}{c}{2.37\%}\\

\multicolumn{1}{c|}{} &

\multicolumn{1}{c|}{MRR}     &
\multicolumn{1}{c}{0.4858}       & 
\multicolumn{1}{c}{0.4300}       &
\multicolumn{1}{c|}{0.5034}    & 
\multicolumn{1}{c|}{0.5055}    & 
\multicolumn{1}{c}{0.5231}      & 
\multicolumn{1}{c}{0.4721}    & 
\multicolumn{1}{c|}{0.5355}      & 
\multicolumn{1}{c|}{\textbf{0.5464}}      & 
\multicolumn{1}{c}{2.04\%}\\ \hline
\end{tabular}
\end{table*}

\subsubsection{Baselines}
Some representative sequential recommendation models are considered as the slow component in the experiments:
\begin{itemize}[leftmargin=*]
	\item[o] \textbf{Caser}~\cite{TangW18}. Caser is a method that combines CNNs and a latent factor model to learn users' sequential and general representations.
	
	\item[o] \textbf{SASRec}~\cite{KangM18}. SASRec is a well-performed model that heavily relies on self-attention mechanisms to identify important items from a user's behavior history. These important items affect user representations and finally determine the next-item prediction.
	
	\item[o] \textbf{DIN}~\cite{zhou2018deep}. DIN is a popular attention model that captures relative interests to target item and obtain adaptive interest representations.
\end{itemize}

We also take the following three well-known lightweight recommendation models as the fast component:
\begin{itemize}[leftmargin=*]
	\item[o] \textbf{FM}~\cite{rendle2010factorization}. This is a benchmark factorization model considering the second-order feature interactions between inputs. Here we treat the IDs of a user and an item as input features.
	
	\item[o] \textbf{NeuMF}~\cite{HeLZNHC17}. This is a pioneering model that combines deep learning with collaborative filtering for general recommendation.
	
	\item[o] \textbf{GRU4Rec}~\cite{hidasi2015session}.
    This is a pioneering model that successfully applies recurrent neural networks to model user sequence for recommendation.
\end{itemize}

To validate the effectiveness of the slow component as the prior of the fast component, we distribute ranking list of candidate items from the slow component to the fast component. As such, the fast component produces more accurate recommendation results based on ranking features from the slow component. Owing to the best performance validated in Table~\ref{tbl:performance-comp} and~\ref{tbl:slow}, we choose DIN as the slow component. Thus, the ad-hoc combination with the fast component can be termed as \textbf{DIN+FM}, \textbf{DIN+NeuMF} and \textbf{DIN+GRU4Rec}.

\begin{table*}[!t]
\centering
\caption{The performance of different models for the slow component.}\label{tbl:slow}
\begin{tabular}{l|ccc|ccc|ccc} 
\hline
\multirow{2}*{Method}& \multicolumn{3}{c}{ML-1M}& \multicolumn{3}{c}{Alipay}&\multicolumn{3}{c}{Steam}\\\cline{2-10}
&HR@5 &NDCG@5 &MRR &HR@5 &NDCG@5 &MRR &HR@5 &NDCG@5 &MRR \\
\hline
MC$^2$-SF & \textbf{0.7166} & \textbf{0.6642}  & \textbf{0.6636}  & \textbf{0.7456}  & \textbf{0.6852} & \textbf{0.6906} & \textbf{0.5853} & \textbf{0.5389} & \textbf{0.5464}\\
\hline

SASRec
& 0.6855 & 0.6188  & 0.6166 & 0.7065  & 0.6501 & 0.6601 & 0.5419 & 0.4961 & 0.5028 \\

Caser
& 0.6837 & 0.6181  & 0.6170 & 0.7024  & 0.6500 & 0.6605 & 0.5393 & 0.4920 & 0.4992 \\
\hline
\end{tabular}
\end{table*}

\subsubsection{Model Implementations}
We implement our model by Tensorflow and deploy it on a Linux server with GPUs of Nvidia Tesla V100 (16G memory).
The model is learned in a mini-batch fashion with a size of 256.
Without specification, all methods optimized by Adam keep the default configuration with $\beta_1=0.9$ and $\beta_2=0.999$.
For the adopted Adam optimizer, we set the learning rate to 5e-4 and keep the other hyper-parameters by default.
We add L2 regularization to the loss function by setting the regularization weight to 1e-4.
The embedding size of all the relevant models is fixed to 32 for ensuring fairness.
The number of layers used in MLP is set to 3.
For reducing the impact of noise, all results in our experiments are averaged over 3 runs.

\subsection{Experimental Results} 
This section elaborates on the comprehensive experimental results to answer the aforementioned three research questions.

\begin{figure*}[!ht]
    \centering
	\includegraphics[width=1.\linewidth]{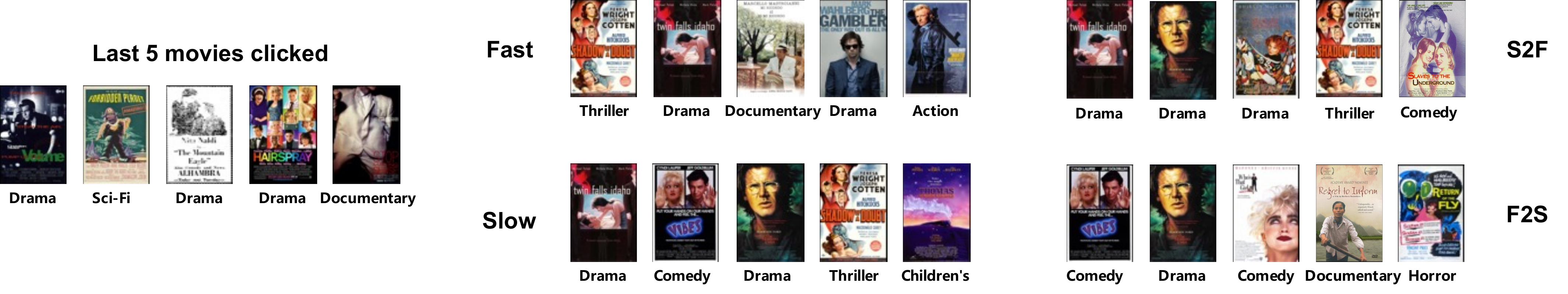}
    \caption{The case study to compare the independent or interactive slow and fast components for one user. The left side presents five items that user clicked recently and the right side gives the top-5 recommendation from four methods. (``Fast'' and ``Slow'' denote independent components, ``S2F'' and ``F2S'' denote interactive components.)}
    \label{fig:case}
\end{figure*}

\subsubsection{Comparison (\textbf{\texttt{Q1}})}
Table~\ref{tbl:performance-comp} presents the overall performance of our model and all the adopted baselines on the mobile side, from which we have the following observations:
\begin{itemize}[leftmargin=*]
\item FM and NeuMF are comparable on each dataset. These two models both focus on feature interaction and user-item collaborative information. Such distinction might be attributed to the characteristics of corresponding datasets. 

\item Compared with other baseline models of the fast component, GRU4Rec achieves the best results on the three datasets.
This conforms to the expectation since only using the representations from the user and item is insufficient, which ignores sequential temporal patterns.
Standard RNNs are good at modeling sequential dependencies, and thus user preference information hidden in behavior sequences could be effectively captured. 

\item Compared to basic fast components that do not consider the guidance of the slow components, all fast components with the privileged information from DIN achieve better performance than the independent counterparts.

\end{itemize}

Table~\ref{tbl:performance-comp} and~\ref{tbl:slow} presents the overall performance of our model and all the adopetd baselines on the cloud side.
\begin{itemize}[leftmargin=*]
\item From the table, Caser achieves poor results on the three datasets. The reason might be that the convolution filter is not good at capturing sequential patterns.

\item SASRec is a transformer-like recommendation model. Due to the strong power of attention computation used in transformer, they perform significantly better than the aforementioned models. However, DIN that models the interaction between the target item and the behavior sequence achieves the best results.
\end{itemize}

\begin{figure}[!t]
    \centering
    \subfloat
    {\includegraphics[width=0.33\linewidth]{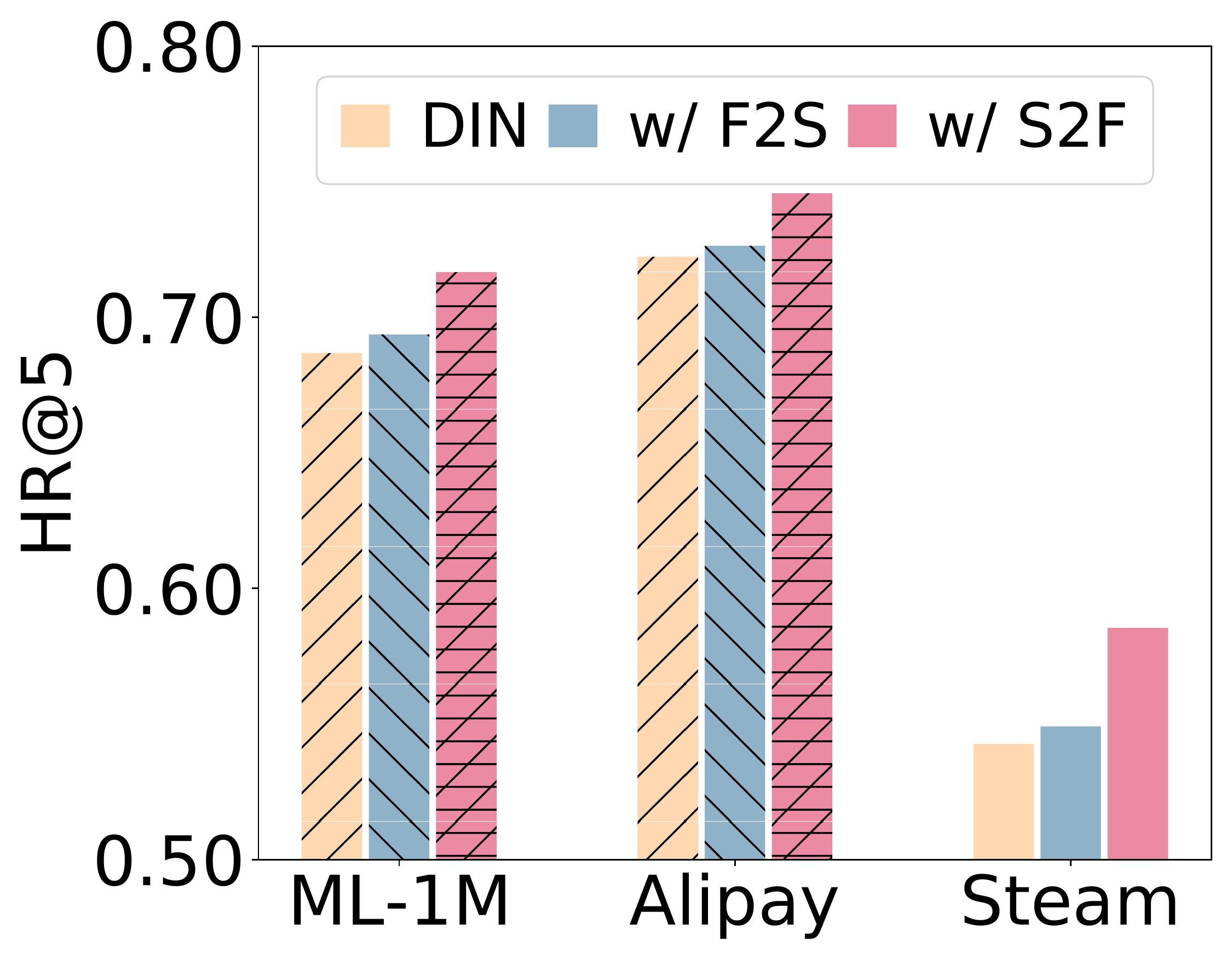}\label{fig:abl_hr}
    \includegraphics[width=0.33\linewidth]{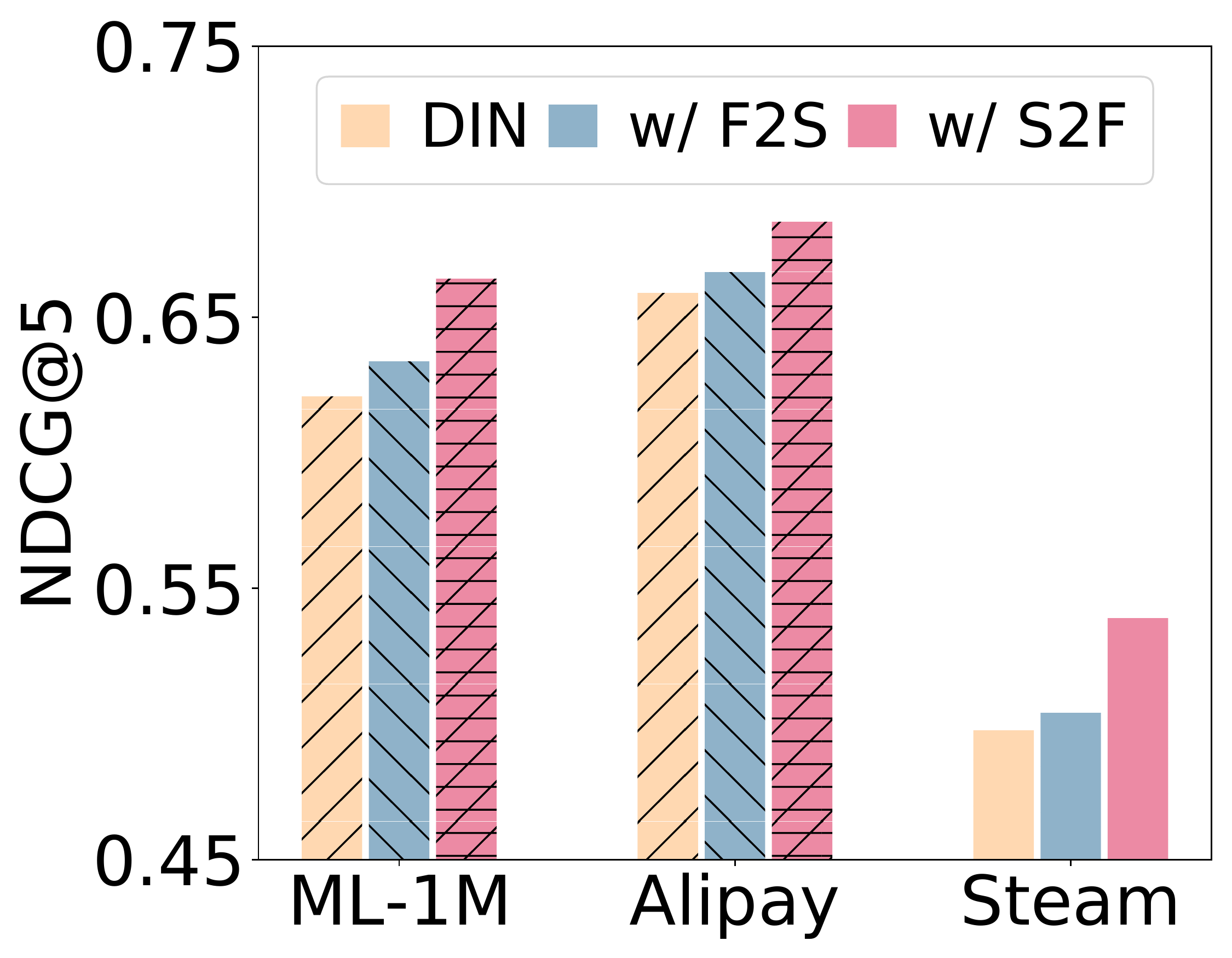}\label{fig:abl_ndcg}
    \includegraphics[width=0.33\linewidth]{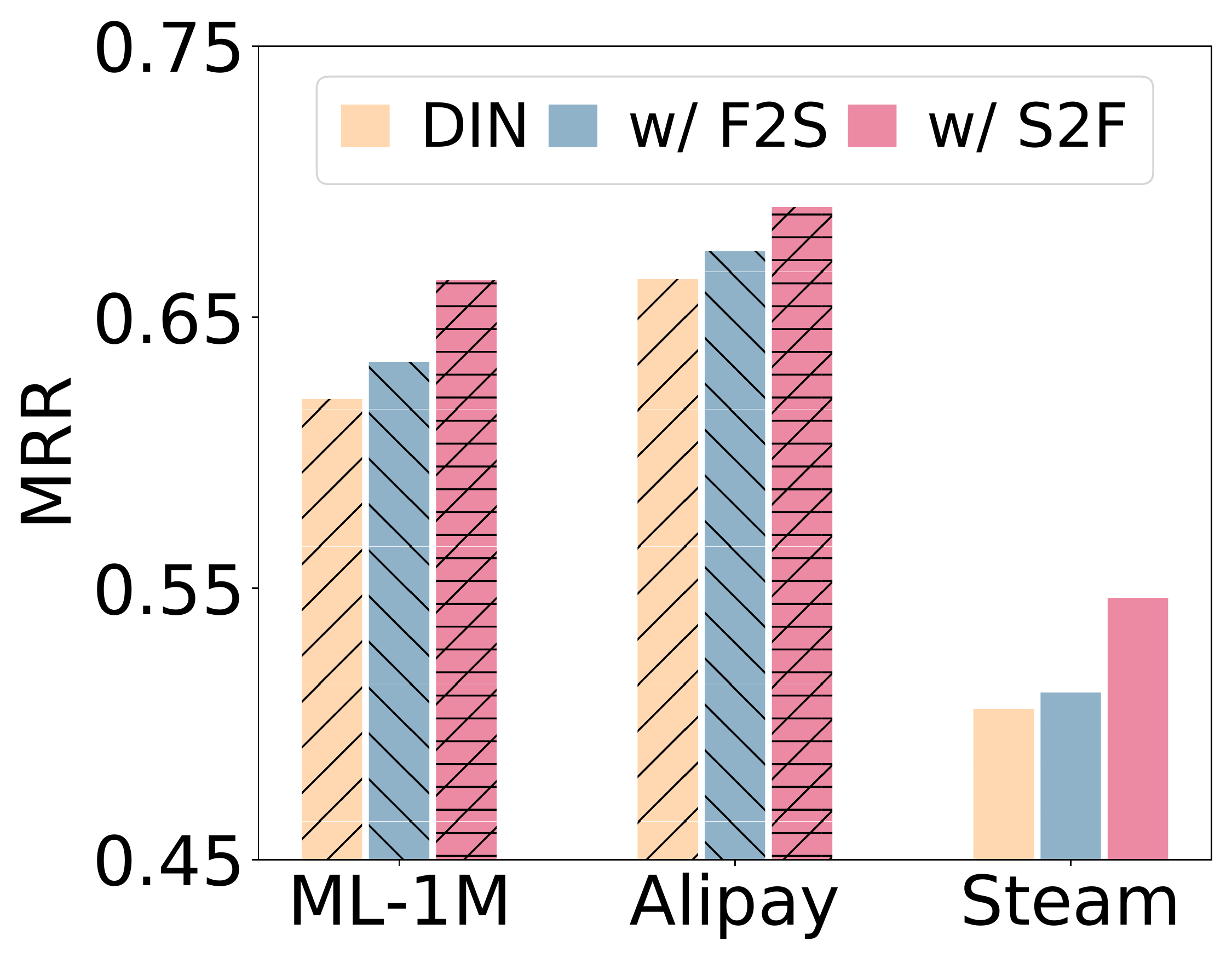}\label{fig:abl_mrr}
    }
    \caption{The performance of MC$^2$-SF compared to DIN.}
    \label{fig:abl}
\end{figure}

As shown in Table~\ref{tbl:performance-comp} and~\ref{tbl:slow}, MC$^2$-SF achieves consistently better performance than all the baselines.
In particular, MC$^2$-SF improves the second-best performed models w.r.t. NDCG@5 by 5.70\%, 1.60\%, and 2.30\% on ML-1M, Alipay, and Steam, respectively. 
This is because: (1) MC$^2$-SF can distill collaborative knowledge from the slow component to the fast component through generalized interest representations. 
(2) The introduction of the negative memory can significantly improve the collaborative recommendation task, which is validated in Figure~\ref{fig:abl}.

\subsubsection{Ablation Study (\textbf{\texttt{Q2}})}
We further conduct the ablation study to validate the contributions of the key components in MC$^2$-SF.
Specifically, (1) ``DIN'' represents using DIN to produce prediction results from perspectives of the slow component. (2) ``w/ F2S'' means uploading negative memory from the fast component to the slow component, so that the slow component could combine these features to refine the predictions. 
(3) ``w/ S2F'' denotes that the slow component distributes privileged features to the fast component on the basis of ``w/ F2S'' and proceeding the collaborative recommendation.

Throughout the result analysis of the ablation study shown in Figure~\ref{fig:abl}, we observe that:

\begin{itemize}[leftmargin=*]
\item ``w/ F2S'' achieves better performance improvement.
It validates the crucial role of negative memory of the fast component.
This is because learning from negative features could lead to more informative user representations which contain positive and negative information.

\item ``w/ S2F'' also shows the significant improvements, which reveals that distributing privileged features and proceeding collaborative recommendation is indispensable. The reason might be that the privileged features could introduce the long-term user interests that the fast component does not have. Meanwhile, collaborative learning for both components is easy to reach equilibrium of the two sides.
\end{itemize}

\begin{figure}[!ht]
    \centering
    \subfloat
    {\includegraphics[width=0.5\linewidth]{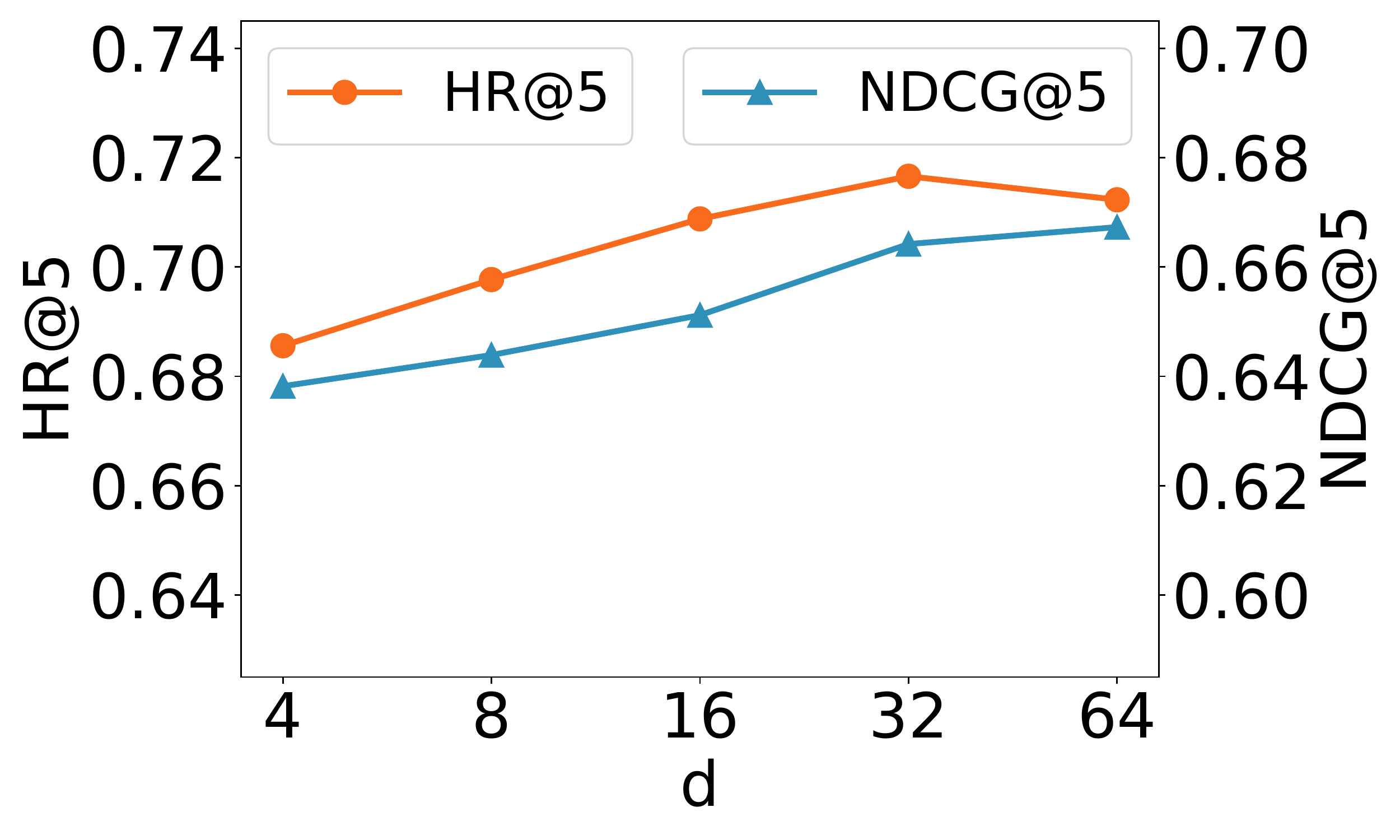}\label{fig:dim}
    \includegraphics[width=0.5\linewidth]{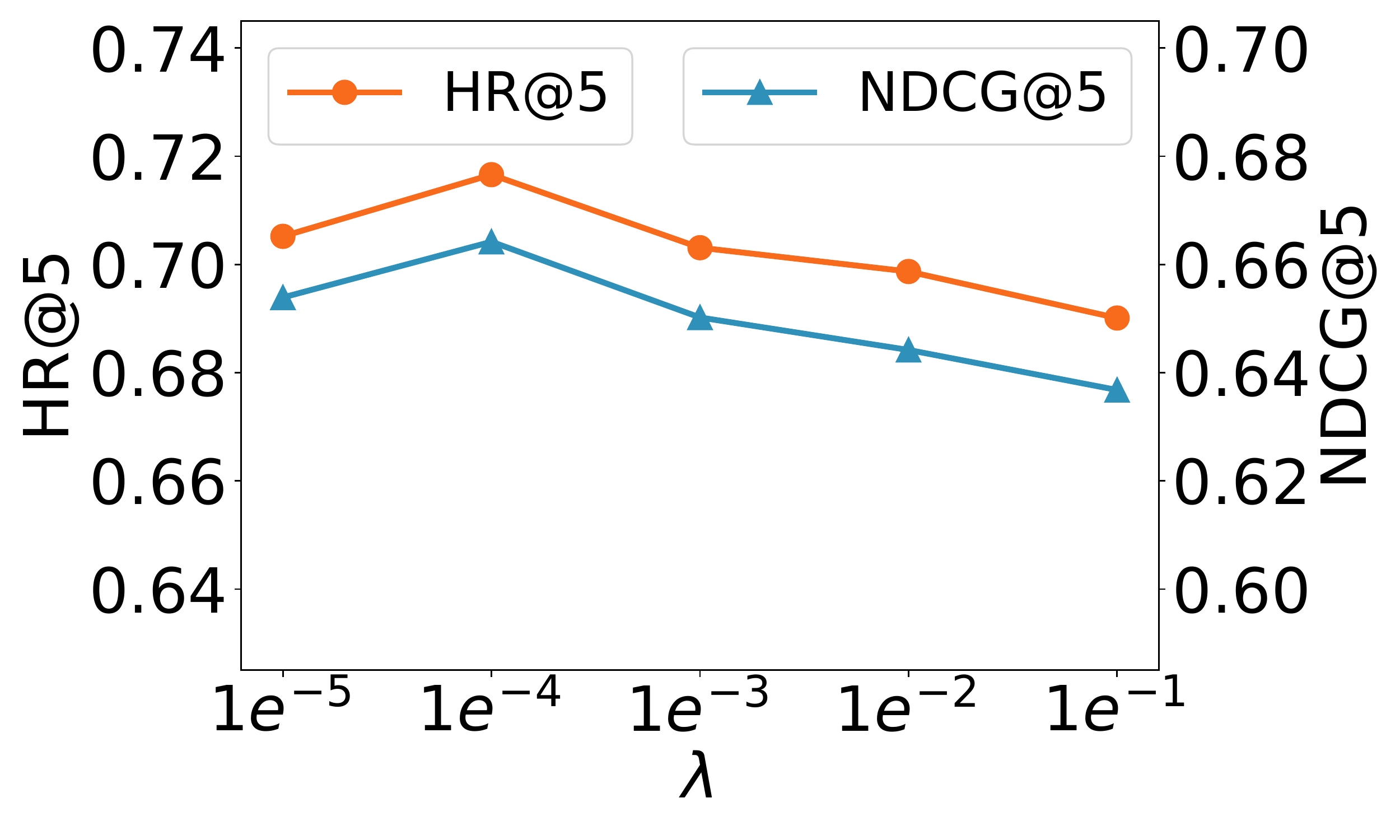}\label{fig:l2}
    }
    \caption{The performance of MC$^2$-SF on ML-1M dataset w.r.t different hyper-parameters.}
    \label{fig:para}
\end{figure}

\subsubsection{Parameter Sensitivity (\textbf{\texttt{Q3}})}
We here study the performance variation for MC$^2$-SF w.r.t. hyper-parameters, and comparative results are shown in Figure~\ref{fig:para}.

\noindent\textbf{Impact of Embedding Size.} We set the embedding size in $\{4,~8,~16,~32,~64\}$. According to the results, the small embedding size might limit the capacity of the model, and too large embedding size is also hard to learn. Only the proper dimension achieves the best performance. 

\noindent\textbf{Impact of Regularization.} We vary the regularization coefficient $\lambda$ among $\{1e^{-5},~1e^{-4},~1e^{-3},~1e^{-2},1e^{-1}\}$, and the best performance is achieved when it is set to $1e^{-4}$. The regularization coefficient $\lambda$ is critical to avoid over-fitting. Too small $\lambda$ cannot constrain the model effectively and too large $\lambda$ may lead to the under-fitting situation.

\subsubsection{Case Study}
To visualize how the collaborative recommendation works, we present the recommendation results of four methods, and illustrate in Figure~\ref{fig:case}.
According to Figure~\ref{fig:case}, (1) ``Fast'' prefers to the prediction relevant to recent click sequence. ``Slow'' recommends based on the sequence in the cloud server, thus the category of movies is somewhat different compared to the five items that user clicked recently.
(2) ``S2F'' mixes the results from ``Fast'' and ``Slow'' to some extent, which considers both the long-term interests and the short-term interests. Besides, the category of children's does not exist in the recommendation results of ``F2S'', which is more relevant to the sequence in the cloud server.

\section{Conclusion}\label{sec:conclusion}
This paper studies mutual benefits of the slow component and the fast component by a Slow-Fast Collaborative Learning framework.
The proposed MC$^2$-SF explores to transfer the prior/privileged knowledge from one side to the other side, and introduces a bidirectional collaborative learning framework to benefit each other. Especially, this work firstly introduces slow-fast ideology to recommendation, resembling the role of System I and System II in the human cognition. The comprehensive experiments conducted on two public datasets and one industrial datasets show the promise of MC$^2$-SF and the effectiveness of its main components.


\bibliographystyle{ACM-Reference-Format}
\bibliography{sample-base}

\appendix

\end{document}